\documentstyle[preprint,aps]{revtex}
\tightenlines
\begin{document} 
\draft

\title{
\begin{flushright}{\small IFT-P.020/98\,\, hep-th/9803219}
\end{flushright}
Note on  the point--splitting procedure to evaluate 
       vacuum fluctuation in certain cylindrically symmetric 
       backgrounds}


\author{I. Brevik\footnote{e-mail: Iver.H.Brevik@mtf.ntnu.no}}
\address{Department of Applied Mechanics\\
         Norwegian University of Science and Technology\\
         N--7034 Trondheim, Norway }
\author{ G.E.A. Matsas \footnote{e-mail: matsas@axp.ift.unesp.br}
and E.S. Moreira Jr. \footnote{e-mail: moreira@axp.ift.unesp.br}}
\address{Instituto de F\'\i sica Te\'orica, 
         Universidade Estadual Paulista,\\
         Rua Pamplona 145, 01405-900, S\~ao Paulo, S\~ao Paulo, Brazil}
\date{\today}
\maketitle


\begin{abstract} 

We revisit two--point function approaches used to study
vacuum fluctuation in  
wedge--shaped regions and conical backgrounds.
Appearance of divergent integrals is discussed 
and circumvented. 
The issue is considered 
in the context of a massless scalar field
in cosmic string spacetime.    

\end{abstract} 
\pacs{04.70.Dy, 04.62.+v}

\narrowtext 

\section{Introduction}

Vacuum fluctuation is one of the most exquisite features of quantum
field theory and has been extensively studied since Casimir's 
seminal work \cite{C}.
Two--point functions are particularly useful to analyze vacuum properties
\cite{BD,F}.  
The Hadamard function
$G^{(1)} (x,x') := \langle \{ \phi(x) , \phi(x') \} \rangle$
gives a comprehensive insight  into  such
fluctuations. For example 
$\langle\phi^2 \rangle$ can be expressed as  
\begin{equation} 
\langle 0|  \phi^2 (x) |0 \rangle = 
                      \lim_{x' \to x} \frac{1}{2} G^{(1)} (x,x') .
\label{Hadamard}
\end{equation}
In some cases, however, it may be preferable to cast the formula above 
in terms of the Feynman propagator, 
\begin{equation}
G_{\cal F} (x,x') = -\frac{i}{2} G^{(1)}(x,x')- \bar G (x,x'),
\label{Fpropagator}
\end{equation}
 as
\begin{equation} 
\langle 0|\phi^2 (x)|0 \rangle = i \lim_{x' \to x} G_{\cal F} (x,x') 
                             + i \lim_{x' \to x} \bar G (x,x') ,
\label{Feynman}
\end{equation}
where 
$\bar G (x,x')$ is the
average of the retarded and advanced Green functions. 
In particular $\bar G (x,x')$  vanishes if $x$ and $x'$
are not causally connected \cite{B}, in which case
Eq. (\ref{Feynman}) becomes
\begin{equation} 
\langle 0 |\phi^2 (x)|0 \rangle = i \lim_{x' \to x} G_{\cal F} (x,x')  .                           
\label{main}
\end{equation}

The present paper was triggered by the following mathematical 
problem. In the process of evaluating Eq. (\ref{main}) in 
a variety of different situations, Deutsch and 
Candelas \cite{DC}, 
Helliwell and Konkowski \cite{HK} and  Matsas \cite{Ma} have made
use of the following  integration formula
\begin{equation}
\int_0^{\infty} dp\ p^{1-\epsilon} J^2_\nu (pr) 
= \frac{2^{1-\epsilon} \Gamma(-1+\epsilon) \Gamma(\nu +1-\epsilon/2)}{
        r^{2-\epsilon} \Gamma^2(\epsilon/2) \Gamma(\nu + \epsilon/2)}
\label{formula}
\end{equation}  
in the $\epsilon \to 0$ limit.
However, as noted in Ref. \cite{BL}  the integral above 
is convergent only when $\epsilon >1$, i.e. Eq. (\ref{formula}) is not valid
in the region  $\epsilon \to 0$. Quite analogously,
an equivalent divergent integral (see Eq.(2.15.3--3) of Ref.\cite{PBM})
was used in a similar context (see Eq. (17) of Ref.\cite{Mo} for $N\geq 3$).
Rather surprisingly, however, a pretty different approach \cite{BL} based 
on Schwinger's source theory led to the same final results
of Refs. \cite{DC,HK,Ma,Mo}. 
As expressed in Sec. VI.C of Ref. \cite{BL},
``We have not got any explanation for the fact that although Deutsch
and Candelas use (\ref{formula}) their final answer is correct.''.

In this paper we revisit the methods used in \cite{DC,HK,Ma,Mo}
giving a more detailed explanation for this delicate behavior.
For sake of clarity we focus the discussion on vacuum fluctuation 
outside cosmic strings but our conclusions must be equally valid
for other cylindrically symmetric backgrounds where the same problem 
appears. In Sec. \ref{sec:Analytic...} 
we discuss the issue using  dimensional analytic 
continuation. In Sec. \ref{sec:Time...} we spot where regularization
takes place.
In Sec. \ref{sec:Conclusions} few remarks are considered
and our conclusions are summarized.

\section{dimensional analytic continuation}
\label{sec:Analytic...}

During past years there has been considerable interest in the study
of vacuum fluctuation outside cosmic strings 
(see \cite{HK,Ma,BL,Mo,general1}  
and references therein). The line element of the spacetime  
outside a cosmic string in $N$ dimensions can be written in the form
\begin{equation}
ds^2 = dt^2 -dr^2 - r^2 d\theta^2 -d{\bf z}^2
\label{le}
\end{equation}
where $0 \leq \theta < \alpha:=2\pi |1-4\mu |$ 
and ${\bf z} \equiv (z_1,...z_{N-3})$.
For $N=4$, $\mu/G$ is the mass per unit 
length of the string. We are assuming an $N$--dimensional spacetime
as this is the parameter which will be used to analytic continue the Green
function into four dimensions.

Considering identification $\theta\sim \theta +\alpha$
the Feynman propagator of a massless scalar field
in this background
can be written for $r = r'$ and ${\bf z} = {\bf z}'$  as \cite{Mo}
\begin{eqnarray}
D_{{\cal F}}^{(N, \alpha)}(\tau; \Delta) 
&=&
\frac{-i}{\alpha} \int_{-\infty}^{+\infty} \frac{d\omega}{2\pi}
\int  \frac{d^{N-3} {\bf k}}{(2\pi )^{N-3}}
\int_{0}^{\infty}dp\ p
\int_{0}^{\infty}ds\  e^{- is(p^2+{\bf k}^2-\omega^2)} e^{-i \omega \tau}
\nonumber \\
&&\times
\sum_{n=-\infty}^{\infty}J^2_{2\pi |n|/\alpha}(p r) 
e^{i 2\pi n \Delta/\alpha} ,
\label{pro2}
\end{eqnarray}
where $\Delta:= \theta-\theta'$ and $\tau:=t-t'$.
By setting $\tau=0$
(note that for $N=4$ one recovers the Feynman propagator 
in \cite{HK}),
making transformations $\omega \to -i\omega$ and $s \to is$
followed by a Wick rotation 
back to the real axes, we  
perform the integrals over $\omega $, $ {\bf k}$ and $s$
obtaining for $N<4$
(see e.g. Ref. \cite{PBM})
\begin{equation}
D_{{\cal F}}^{(N, \alpha)}(\Delta) =
\frac{-i \pi^{1-N/2}\Gamma(2-N/2)}{2^{N-2} \alpha}
\sum_{n=-\infty}^{\infty} e^{i 2\pi n \Delta/\alpha}
\int_{0}^{\infty} dp\ p^{N-3}
J^2_{2\pi |n|/\alpha}(p r) .
\label{GFN2}
\end{equation}
For $N=4$ the integral that appears above is the divergent one
in Eq. (\ref{formula}). Notwithstanding
it is convergent in the interval $2<N<3$ leading to 
\begin{eqnarray}
D_{{\cal F}}^{(N, \alpha)}(\Delta) 
 & = & 
\frac{-i\Gamma[(3-N)/2]}{ \alpha(2 \sqrt{\pi})^{N-1}  r^{N-2} }
\sum_{n=-\infty}^{\infty} 
{
\frac{\Gamma(-1+N/2 + 2\pi |n|/\alpha) }{\Gamma(2 - N/2 +2\pi |n|/ \alpha)}
e^{i 2 \pi n \Delta/\alpha} 
}
\label{GFN3}
\end{eqnarray}
in agreement with Ref. \cite{Mo}.
Assuming now that  
both sides of Eq. (\ref{GFN3})
are analytic functions of $N$, by the principle of analytic continuation
equality (\ref{GFN3}) also holds for $N\geq 3$.
Thus after performing the
summation with $N=4$ one obtains
\begin{equation}
D_{{\cal F}}^{(4, \alpha)}(\Delta) =
-\frac{i}{4 \alpha^2 r^{2}}
\csc ^{2}\frac{\pi \Delta}{\alpha} ,
\label{final1}
\end{equation}
which is the correct result as has been 
rigorously shown in Ref. \cite{BL}.

The reason why we (and the authors of Refs. \cite{DC,HK,Ma,Mo})
have arrived to the correct result
may be traced back 
to Eq. (\ref{GFN2}). Namely 
the divergent
integral in the r.h.s. of  Eq. (\ref{GFN2}) is regularized 
by the summation. Thus although the integral itself is not analytic
when $N=4$ since it is divergent, the r.h.s. of
Eq. (\ref{GFN2}) as a whole is analytic.  In the next section we will
show how the divergence can be isolated and explicitly 
regularized.

\section{Regularization procedure}
\label{sec:Time...}

The regularization procedure adopted in this section will
be possible by considering in Eq. (\ref{pro2}) $\tau \neq 0$
(eventually the limit $\tau\rightarrow 0$ will be taken)
and that the interval associated with $(x,x')$ is spacelike
[then vacuum expectation values can be extracted from the
Feynman propagator as in Eq. (\ref{main})].
This can be implemented by using 
an approach slightly different from the one above.

Performing 
integrations in Eq. (\ref{pro2}) except over the $s$ variable we obtain
\cite{Mo}
\begin{equation}
D_{{\cal F}}^{(N, \alpha)}(\tau; \Delta)=
\frac{2\pi }{\alpha} \int_{0}^{\infty}\frac{ds}
{(4\pi i s)^{N/2}}
e^{-i(\tau^{2}-2 r^{2})/4s}
\sum_{n=-\infty}^{\infty}I_{2\pi |n|/\alpha}
\left(r^{2}/2is \right)e^{i2\pi n\Delta/\alpha} .
\label{pro3}
\end{equation}
The remaining integral above can be evaluated by 
first adding an infinitesimal negative imaginary part
to $\tau^{2}$ and then by using \cite{PBM}
\begin{equation}
\int_{0}^{\infty}dx\  x^{\mu-1}e^{-px}I_{\nu}(cx) =
p^{-(\mu+\nu)}\left(\frac{c}{2}\right)^{\nu}
\frac{\Gamma[\nu+\mu]}
{\Gamma[\nu+1]}
{}_{2}F_{1}\left[\frac{\mu+\nu}{2}, \frac{\mu+\nu+1}{2};
\nu+1;\frac{c^{2}}{p^{2}}\right],
\label{formula1}
\end{equation}
which holds for 
${\rm Re}(\mu+\nu)>0$ and ${\rm Re}\ p>|{\rm Re}\ c|$.
The hypergeometric function in Eq. (\ref{formula1}) can be recast
by considering a linear transformation (see e.g. Ref. \cite{GR}),
\begin{eqnarray}
{}_{2}F_{1}\left[a,b;c;z\right] &=&
\frac{\Gamma[c]\Gamma[c-a-b]}
{\Gamma[c-a]\Gamma[c-b]}
\ {}_{2}F_{1}\left[a,b;1+a+b-c;1-z\right]
\nonumber
\\ 
&+& 
\frac{\Gamma[c]\Gamma[a+b-c]}
{\Gamma[a]\Gamma[b]}
\ {}_{2}F_{1}\left[c-a,c-b;1-a-b+c;1-z\right] (1-z)^{c-a-b},
\nonumber
\end{eqnarray}
whose domain of validity is the $z$-plane cut along
the real axis from $z=-\infty$ to $z=0$
and from $z=1$ to $z=\infty$; with 
\begin{equation}
a+b-c\neq 0,\pm 1,\pm 2, \ldots  
\label{nonodd}
\end{equation}

Thus the Feynman propagator can be written as shown in Eq. 
(\ref{Fpropagator}) with
\begin{eqnarray}
{D^{(1)}}^{(N, \alpha )} (\tau; \Delta ) 
             &=& \frac{1}{\alpha} \sum_{n=-\infty}^{\infty} e^{i2\pi n \Delta/\alpha}
             \frac{\pi^{(1-N)/2} r^{2-N}}{2^{N - 2}}
             \left( 1-\frac{\tau^2}{2r^2}\right)^{1-N/2 - 2\pi |n|/\alpha}
             \nonumber \\
             &\times &  
             \frac{\Gamma[-1+N/2 + 2\pi |n| /\alpha] \Gamma[(3-N)/2]}
             {\Gamma[2 -N/2+2\pi |n| /\alpha]} 
             \nonumber \\
             &\times &  
             {}_{2}F_{1}
             \left[
             \frac{\pi |n|}{\alpha} + \frac{N-2}{4},
             \frac{\pi |n|}{\alpha} + \frac{N}{4};
             \frac{N-1}{2};
             1-\left(1-\frac{\tau^2}{2r^2}\right)^{-2}
             \right]
\nonumber
\end{eqnarray}
and
\begin{eqnarray}
{\bar D}^{(N, \alpha )} (\tau; \Delta )
             &=& \frac{i}{4 \alpha} \sum_{n=-\infty}^{\infty} e^{i2\pi n \Delta/\alpha}
             \pi^{(1-N)/2}r^{2-N}
             \left(1-\frac{\tau^2}{2r^2}\right)^{1-N/2 - 2\pi |n|/\alpha}
             \nonumber \\
             &\times &
             \left[1-\left(1-\frac{\tau^2}{2r^2}\right)^{-2}\right]^{(3-N)/2}
             \Gamma\left( \frac{N-3}{2} \right)
             \nonumber \\
             &\times &  
             {}_{2}F_{1}
             \left[
             \frac{|n| \pi}{\alpha } + \frac{6-N}{4},
             \frac{|n| \pi}{\alpha} + \frac{4-N}{4};
             \frac{5-N}{2};
             1-\left(1-\frac{\tau^2}{2r^2} \right)^{-2}
             \right].
\nonumber
\end{eqnarray}
Using these expressions for small $\tau$, the Feynman propagator
becomes 
\begin{eqnarray}
D_{{\cal F}}^{(N, \alpha)}(\tau; \Delta)&=&
\frac{ -i\Gamma [(3-N)/2]}{\alpha (2 \sqrt{\pi})^{N-1} r^{N-2}}
\sum_{n=-\infty}^{\infty} \frac{\Gamma (-1+N/2 +2\pi |n|/\alpha)}{\Gamma (2-N/2 +2\pi |n|/\alpha)} 
e^{i2\pi n \Delta/\alpha }
\nonumber \\
&&-\frac{i^{-N} \Gamma[(N-3)/2]}{4\pi^{(N-1)/2} r |\tau|^{N-3}}
\sum_{n=-\infty}^{\infty} \delta(\Delta +\alpha n)
\label{final2}
\end{eqnarray}
which holds for 
non odd $N$ [as follows from Eq. (\ref{nonodd})],
$N>2$. 
In obtaining Eq. (\ref{final2}) we have also used
Poisson's Formula,
\begin{displaymath}
\sum_{m=-\infty}^{\infty}\delta(\phi +2\pi m)=
\frac{1}{2\pi}\sum_{m=-\infty}^{\infty}\exp\left\{im\phi\right\}.
\end{displaymath}

By taking $\tau\rightarrow 0$ in Eq. (\ref{final2})
when $N<3$ we recover Eq. (\ref{GFN3}) as should be. Now when $N > 3$
it may appear that we are left with a divergence as $\tau \to 0$
due to the factor $1/|\tau|^{N-3}$ 
in the second term on the r.h.s. of Eq. (\ref{final2}).
However, before taking $\tau \to 0$, we should note that 
$\delta(\Delta +\alpha n)$
vanishes identically for every $n$ since $\Delta \neq \alpha n$
(recall that we are dealing with a spacelike interval). Thus
the second term in the r.h.s. of Eq. (\ref{final2}) is null
and Eq. (\ref{GFN3}) still holds for $N>3$.

\section{Final Remarks}
\label{sec:Conclusions}

Before closing the paper let us discuss a couple of points.
As mentioned above, Eq. (\ref{GFN3}) is not suitable
for odd $N$.
For example, when $N=3$, $\Gamma(0)\delta(\Delta)$
arises in Eq. (\ref{GFN3}) which needs to be regularized. 
Using $\Gamma[a+z]/\Gamma[b+z]=
B[z+a,b-a]/\Gamma[b-a]$ with $B[z,w]$ denoting the Euler beta function,
we see that Eq. (\ref{GFN3}) may be rewritten as
\begin{eqnarray}
D_{{\cal F}}^{(N,\alpha)}(\Delta)&=&
\frac{1}
{i\alpha 2\pi^{N/2-1}r^{N-2}       
\Gamma\left[(4-N)/2\right]}
\label{mdell}
\\
&&\times\left\{
B\left[\frac{N-2}{2},3-N\right]
+2\sum_{m=1}^{\infty}
B\left[\frac{2\pi m}{\alpha}+\frac{N-2}{2},3-N\right]
\cos \frac{2\pi m\Delta}{\alpha}\right\},
\nonumber
\end{eqnarray}
where we have used the fact that 
$\Gamma[2x]/\Gamma[x]=2^{2x-1}\pi^{-1/2}\Gamma[x+1/2]$.
By using 
$B[z,w]:=\int_{0}^{1}dt\ t^{z-1}(1-t)^{w-1}$ in the 
second term in Eq. (\ref{mdell})  
and performing the summation \cite{GR}, we are left with the following 
integral representation \cite{mor97}
\begin{displaymath}
D_{{\cal F}}^{(N,\alpha)}(\Delta)=
\frac{1}
{i\alpha 2\pi^{N/2-1}r^{N-2}       
\Gamma\left[(4-N)/2\right]}
\int_{0}^{1}dt\ t^{(t-4)/2}\frac{(1-t)^{2-N}(1-t^{4\pi /\alpha})}
{1-2t^{2\pi /\alpha}\cos\{2\pi \Delta/\alpha\}+t^{4\pi /\alpha}},
\end{displaymath}
which is suitable for odd $N$ (but not for even $N$).

Finally let us consider a check of consistency.
For $N=4$, which is our main interest here,
one obtains from Eq. (\ref{final2})
\begin{equation}
D_{{\cal F}}^{(4,\alpha)}(\tau; \Delta)=
-\frac{i}{4 \alpha^2 r^{2}}
\csc ^{2}\frac{\pi \Delta}{\alpha} -
\frac{1}{4\pi r |\tau|}
\sum_{n=-\infty}^{\infty}
\delta(\Delta+ \alpha n)
\label{propagator}.
\end{equation}
The first term in Eq. (\ref{propagator}) is the 
familiar expression (\ref{final1}). 
The second term, which vanishes, was kept just for formal
comparison with the usual Minkowski spacetime result. Indeed
Eq. (\ref{propagator}) with $\alpha=2\pi$  must be consistent
with the usual expression in Minkowski spacetime 
\cite{BD}
\begin{equation}
D_{{\cal F}}^{(4,2\pi)}(x,x') =
\frac{i}{8 \pi^2 \sigma} 
-\frac{\delta(\sigma)}{8\pi } 
\label{Mpropagator}
\end{equation} 
where $2\sigma := |x^\mu - {x'}^\mu|^2 $. Starting from
Eq. (\ref{Mpropagator}) this fact can be 
verified by noting that (for $r= r'$ and 
${\bf z}={\bf z'}$) in cylindrical coordinates 
$2\sigma = \tau^2 - 2 r^2 + 2 r^2 \cos \Delta$.
Then using formula
$\delta[f(x)]=\sum_{j}\delta(x-x_{j})/|f'(x_{j})|$,
where $x_{j}$ are the simple zeros of $f(x)$,
it is rather straightforward to reproduce 
Eq. (\ref{propagator}) in the limit of small $\tau$.

In summary we have investigated the reason why a number of previous works
have obtained correct result for 
vacuum fluctuation in wedge--shaped regions and conical backgrounds
although using illicit expressions for divergent integrals. 
The problem is solved by 
noting that there is an overall summation 
which renders the result 
finite.  In Sec. \ref{sec:Analytic...} we use the principle
of analytic continuation to obtain a definite result, 
while in Sec. \ref{sec:Time...} such principle is verified
by isolating the divergence and regularizing. 
With these clarifications in mind those
works that have obtained correct answers using illegal
intermediate steps will not have their results seen as incidental.

\acknowledgments
GM and EM would like to acknowledge partial and full financial support 
respectively from  Conselho Nacional de Desenvolvimento Cient\'\i fico e
Tecnol\'ogico and  Funda\c c\~ao de Amparo \`a Pesquisa
do Estado de S\~ao Paulo.

\end{document}